\documentstyle[epsf,11pt,amssymb]{article}

\begin{document}
\baselineskip=5mm

\twocolumn[
\begin{center}

\noindent {\LARGE \bf Anomalies in Universal Intensity Scaling in}

\

\noindent {\LARGE \bf Ultra-relativistic Laser-Plasma Interactions}

\vspace*{7mm}

\noindent {\ T.J.M.\ Boyd$^{1}$ and R.\ Ondarza-Rovira$^{2}$ }

\

\noindent $^{1}${\it Centre for Theoretical Physics, University of Essex, Wivenhoe~Park, Colchester~CO4~3SQ, UK}

\vspace*{0.7mm}

\noindent $^{2}${\it Instituto Nacional de Investigaciones Nucleares, A.P.\ 18-1027, M\'{e}xico 11801, Distrito Federal, Mexico} 
\end{center}

\begin{center}
\parbox{14cm}{\small Laser light incident on targets at intensities
such that the electron dynamics is ultra-relativistic, gives rise to
a harmonic power spectrum extending to high orders and characterized by
a relatively slow decay with the harmonic number $m$ that follows a 
power law dependence, $m^{-p}$. Relativistic similarity theory predicts
a {\it universal} value for $p = 8/3$ up to some cut-off $m = m^{*}$. 
Results presented in this Letter suggest that under conditions in which
plasma effects contribute to the emission spectrum, the extent of this 
contribution may invalidate the concept of universal decay. We report a 
decay with harmonic number in the ultra-relativistic range 
characterised by an index $5/3 \lesssim p \lesssim 7/3$, 
significantly weaker than that predicted by the similarity model.

\vspace*{3mm}

\noindent PACS numbers: 52.38.Kd, 42.65.Ky, 52.65.Rr }
\end{center} \vspace*{-0.5cm} ]

Since the earliest observations of harmonics of laser light 
reflected from target plasmas by Burnett {\it et al.\ }[1] and Carman, 
Rhodes and Benjamin [2], the topic of harmonic generation has been one 
of enduring interest. The range of harmonics excited is governed in 
large part by the level of $I \lambda_{L}^{2}$ where $I$ is the 
intensity of the incident light and $\lambda_{L}$ its wavelength. 
Levels of $I\lambda_{L}^{2}$ attained in recent experiments are such 
that the laser-plasma interaction physics extends far into the 
ultra-relativistic (UR) regime. Harmonic orders ranging from 
$\sim 850$ for peak intensities
a little over $I_{20} \equiv 10^{20}\, $Wcm$^{-2}$ to $\sim 3200$ when 
intensities are an order of magnitude greater have been 
reported by Dromey {\it et al.} Ref.\ [3]. In this work the decay of 
intensity with harmonic order is found to behave as
$I_{m} \sim m^{-p}$ where $2.2 \lesssim p \lesssim 2.7$ at intensities 
$I \sim I_{20}$ and $2.5 \lesssim p \lesssim 3.0$ at intensities in 
the higher range.
Harmonic decay with $p$ in the range observed is consistent  
with an intriguing prediction by Baeva, Gordienko and Pukhov (BGP) [4] 
based on
a similarity analysis which led them to conclude that the harmonic 
decay from ultra-relativistic laser-plasma interactions is independent
of the details of the interaction physics and that the value of
the decay index $p = 8/3$ is in fact {\it universal}. 

In this Letter we report results from a set of particle-in-cell (PIC) 
simulations in the UR range which suggest that the decay of harmonic 
intensities may be governed by plasma effects. In these circumstances
we present evidence of spectral decay more complex than the universal
decay proposed by BGP, no longer characterised by a unique index 
$p=8/3$ but by indices spanning a range $5/3 \lesssim p \lesssim 7/3$. 

The effects that appear in the spectrum reflect details of the 
interaction physics, in particular the plasma oscillations generated by 
bursts of highly relativistic electrons. We have shown elsewhere  
that with $p-$polarised light, collective effects not only give rise to
emission at the plasma frequency and its harmonics [5] but can serve as
a source of a modulation at the plasma frequency that appears in the 
spectrum for suitably chosen parameters [6]. 
Other work dealing with the effects of plasma oscillations driven by 
energetic electrons on emission spectra has been reported by Teubner 
{\it et al.\ }[7] and by Eidmann {\it et al.\ }[8].

In the simulations carried out to generate the harmonic spectra we treat
the interaction  of a laser pulse of length $\tau_{p}$ with a plasma 
slab of initially uniform density, apart from an interface at the front
end. This vacuum-plasma interface is characterised by a density profile
with scale length $\Delta$ that is some prescribed fraction of a 
wavelength across the ramp. 
We have used a 1-1/2D fully relativistic and explicit PIC code, that 
embeds the Bourdier technique [9] to allow for oblique incidence.
Two vacuum gaps extend from both the 
front of the ramp and the planar rear surface of the plasma to the 
walls of the simulation box to allow for particle and wave
propagation. When external fields reach the boundary and propagate
outside the system they are not reflected and thus no longer 
considered in the interaction physics. Plasma particles are reflected
at the boundaries by reversing their velocities.
The plasma filled a simulation box extending over 4 laser wavelengths 
with $2 \times 10^{6}$ particles distributed
across $10^{4}$ grids, giving a ratio of grid size to 
wavelength of $\sim 10^{-4}$. The number 
of particles employed and the choice of the simulation parameters is
sufficient to resolve both the Debye length and the highest 
frequency mode with acceptable resolution.

The initial electron temperature was chosen to be 100 eV and we used
gaussian pulses of variable length and profile. Ions were treated
as a neutralizing immobile background, an acceptable approximation 
given the femtosecond time scales governing these simulations where a
17 fs gaussian pulse is used throughout. The normalized quiver 
momentum $a_0 \sim 8.5\, {\left( I_{20} \lambda_L^2 \right)}^{1/2}$ 
lay in the range 5-20 with slab densities between 20 $n_c$ and 200 
$n_c$, where $n_c$ is the critical density. The reflected emission 
spectra showing the relative 
strength of the different oscillation modes are determined from a 
time-resolved Fourier decomposition of the harmonic
contents of the reflected electric field at the vacuum gap, normalized 
to the fundamental.

\renewcommand{\thefigure}{1}
\begin{figure}[h]  
{\bf \epsfxsize=9.0cm \epsfysize=9.0cm \epsffile{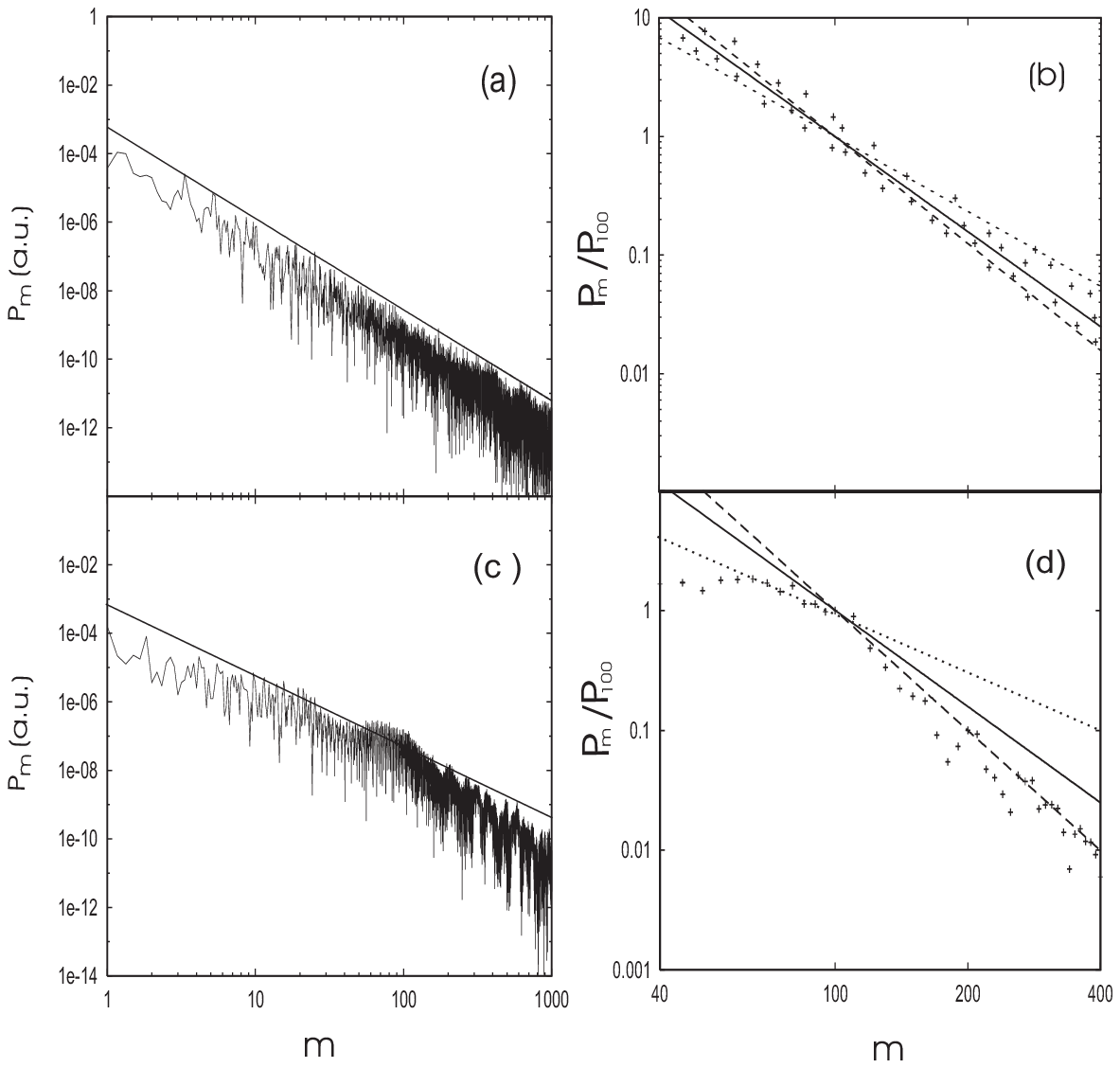} \vspace*{0.5cm}}
\parbox{9cm}{Figure 1: \small Harmonic spectrum generated from a 
laser pulse incident normally (ab) and obliquely (cd) on a plasma 
target: $a_0=10, n_e/n_c=40, \tau_{p} = 17$ fs and
$\lambda_{L} = 1054$ nm. In (a) the line represents spectral decay
$P_{m} \sim m^{-p}$ with a decay index $p = 8/3\,$, 
(b) relative intensity of harmonics in (a) scaled to $m = 100\,$, 
i.e.\ $P_m/P_{100} = (m/100)^{-p}\,$. The lines correspond to 
$p=2.1$ (dotted), $p = 8/3$ (solid) and $p= 3.0$ (dashed);
(c) harmonic spectrum generated by obliquely incident 
laser pulse showing the effects of spectral modulation; the line
denotes a spectral decay $p = 2$;
(d) relative intensity of harmonics in (c) scaled to $m = 100$; 
the lines correspond to 
$p = 5/3$ (dotted), $p = 8/3$ (solid) and $p= 10/3$ (dashed).}
\end{figure}

In Fig.\ 1 we contrast PIC spectra from runs at UR intensities with 
light incident both normally (Fig.\ 1ab) and obliquely (Fig.\ 1cd) 
on the surface of the plasma slab. The plasma density is chosen to be 
initially uniform with $n_e= 40\, n_c$ and the peak intensity of the 
incident pulse such that $a_0 = 10$. Fig.\ 1a shows harmonic power 
levels up to $m = 1000$; harmonics up to $m \sim 200$ are characterised 
by conversion efficiencies $\geq 10^{-6}$. The decay in peak harmonic 
power levels 
with harmonic number $m$ is well represented across the range by the 
index $p = 8/3$, in agreement with both the similarity analysis and the 
PIC spectrum due to Baeva, Gordienko and Pukhov [4]. At the same time 
it appears that across the harmonic range in Fig.\ 1a there is no  
marked break in the spectrum predicted by their model at a 
critical harmonic $m^{*}=\sqrt{8\, \alpha}\, \gamma_{s}^{3}$ where 
$\gamma_{s}$ is the maximum value of the relativistic factor
associated with the motion of the plasma surface and $\alpha$, the 
second derivative of the surface velocity, is of order 1. 
For the parameters in Fig.\ 1, $m^{*} \sim 40 - 80$, 
adopting the BGP estimate for $\gamma_{s}$.

Fig.\ 1b presents data from the spectrum in Fig.\ 1a in which a set of 
data points is represented over the restricted harmonic range 
$m = 40$ to $m = 400$ to afford a close-up of the macroscopic intensity
scaling applied across the full harmonic range in Fig.\ 1a. 
We arbitrarily normalize the harmonic power levels to $P_{100}$ 
and find from Fig.\ 1b that across the harmonic range 
(40 - 400) the scaling with harmonic number falls in a range 
$2.1 \lesssim p \lesssim 3.0$. In passing we note that this result is 
not greatly at variance with the scaling reported in [3] where 
$2.2 \lesssim p \lesssim 2.7$ and $2.5 \lesssim p \lesssim 3.0$ (the
latter applied across a much higher harmonic range),
depending on intensity. We draw no comparison between our simulated
scaling and those from Ref.\ [3] in view of 
differences in parameters 
such as the pulse length (17 fs in the simulations as against 500 
fs in the experiments) and without knowing 
the experimental plasma electron density. 
That notwithstanding, the results for $s$-polarised light in Figs.\ 1a,b
are broadly consonant with the BGP scaling $p = 8/3$. 

Figs.\ 1c, 1d highlight changes in the pattern of harmonic emission 
when $p$-polarised light is incident at $23^{o}$ to target 
normal, these spectra showing distinct differences from those at normal
incidence or for $s$-polar\-ised light. 
It is at once apparent that on a macroscopic level, the spectrum
is no longer well characterized by a universal decay index $p = 8/3$.
In contrast to the spectrum in Fig.\ 1a there 
{\it is} now a more or less well-defined break in the spectrum in
Fig.\ 1c, centred on $m \sim 100$. We emphasize that this feature is 
quite distinct from the spectral break deriving from a validity 
condition in the BGP self-similar model. 
The break in Fig.\ 1c occurs with the onset of modulation; below this 
the spectrum is enhanced over that for $s$-polarised 
light (cf. Fig.\ 1a) 
by plasma emission with the 
consequence that the decay coefficient across this region 
is reduced below the BGP ``universal index''.
We show in Fig.\ 1d an intensity scaling up to $m = 400$ corresponding 
to that in Fig.\ 1b, with harmonic power levels normalised to $P_{100}$,
characteristic of the distinctive modulation in the spectrum 
in Fig.\ 1c. 
Below $m = 100$ decay is less rapid than the corresponding
universal decay, with $p \sim 5/3$, while above this the best
fit for this choice of parameters is given by $p \sim 10/3$. 

\renewcommand{\thefigure}{2}
\begin{figure}[h] 
{\bf \epsfxsize=9cm \epsfysize=9cm \epsffile{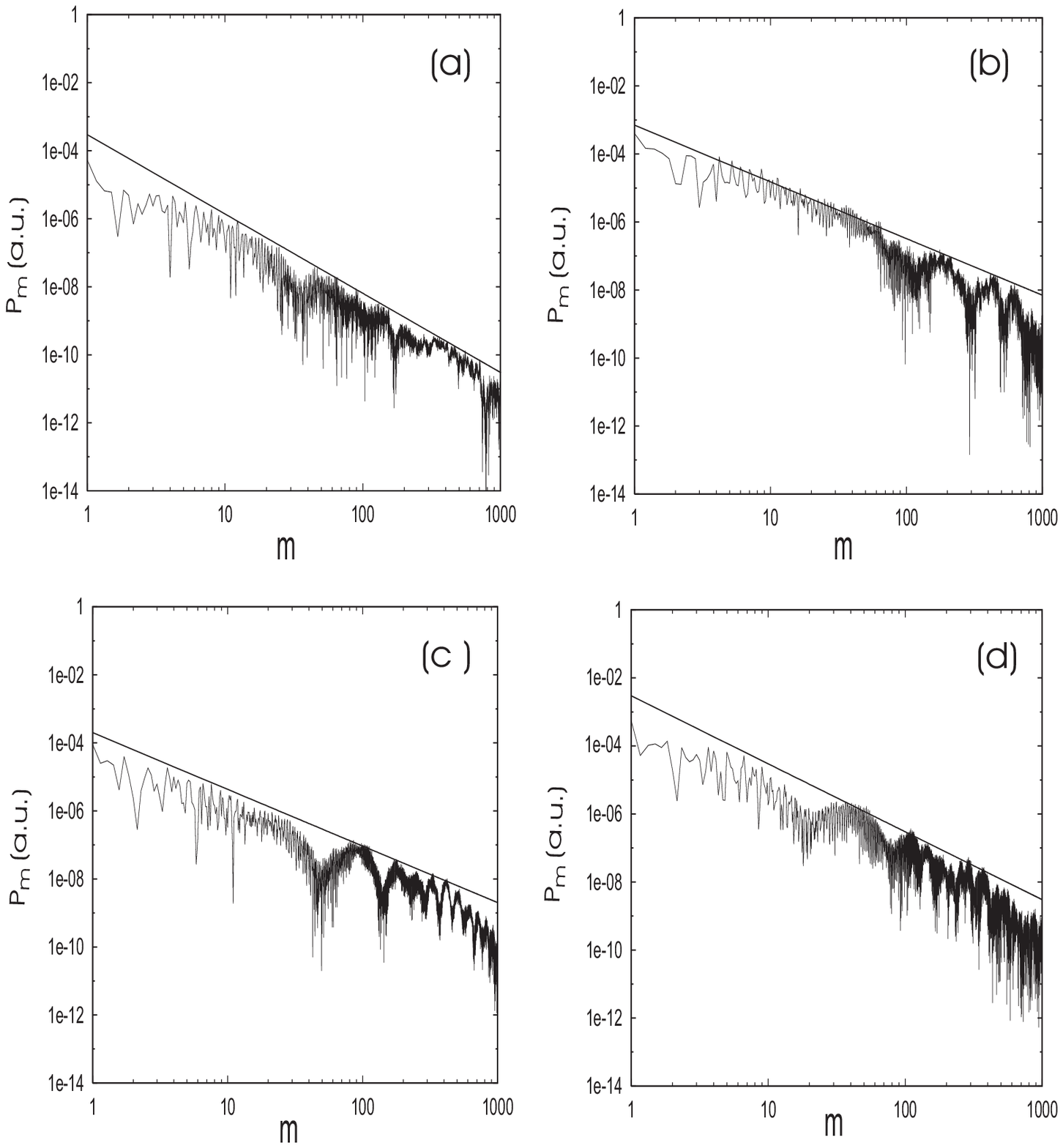} \vspace*{0.5cm}}
\parbox{9cm}{Figure 2: \small Modulation-induced an\-omalies in the 
universal scaling of har\-monic spectra from ultra-relativistic 
laser-plasma interactions. The laser pulse with pulse-length $\tau_p = 
17$ fs, is incident obliquely on a plane target at an angle of incidence
$\theta =23^{o}$ and the wavelength $\lambda_L = 1054$ nm. (a) $a_0=5, 
n_e/n_c=20$, (b) $a_0=20$, $n_e/n_c=40$, (c) $a_0=10, 
n_e/n_c=50$, (d) $a_0=20, n_e/n_c=100$. The lines indicate a scaling
$P_{m} = m^{-p}$, where $p = 7/3$ in (a), $p = 5/3$ in (b),(c) and 
$p = 2$ in (d).}
\end{figure}

In Fig.\ 2 we present a range of spectra for additional combinations of 
$a_{0}$ and $n_{e}/n_{c}$, namely $(a_{0}, n_{e}/n_{c}) = (5, 20)$,  
$(20, 40)$, $(10, 50)$ and $(20, 100)$. All four combinations show 
spectra with distinctive structure across the range and serve 
to confirm the deviation from BGP universal scaling already 
apparent from Fig.\ 1c. We 
find some variation in the onset of modulation depending on
particular combinations of $a_0$ and $n_e/n_c$, with thresholds ranging 
from $m \sim 20 - 30$ (Fig.\ 2a,d) to $m \sim 50 - 70$ (Fig.\ 2b,c). 
Not only
is there variation in the onset of modulation, there are also notable 
differences in modulation frequency across the range of parameters 
considered. This contrasts with our findings in Ref.\ [6] where, for a
moderately relativistic range, spectra showed fairly regular modulation
at the plasma frequency $\omega_p$, with
onset of modulation consistently at $\sim 1.5\, \omega_p$.
Despite the variation in detail in the spectra in Fig.\ 2, the decay 
index in each case is tolerably well fitted by $p$ in the range 
$5/3 \lesssim p \lesssim 7/3$, markedly less than the spectral decay 
observed in simulations at lower $a_0$ [6]. 

The physics differentiating irradiation by 
{\it p-}polarized from {\it s-}polarized light derives from bursts of 
Brunel-excited
electrons [10] that drive plasma oscillations to highly   
nonthermal levels close to the plasma surface.
Given the steep gradients in density at the front of the
simulation box, the plasma waves excited 
are no longer uniquely longitudinal but couple strongly to 
the radiation 
field, giving rise to emission at the plasma frequency in both 
back and forward directions, with weaker contributions at higher 
harmonics of the plasma frequency [5]. Boyd and Ondarza-Rovira further 
identified a robust feature in the spectrum that
appeared between the plasma line and its second harmonic, typically at 
a frequency $1.5\, \omega_p$, with width of the order of the plasma 
frequency. This so-called `combination line' proved in the event  
to be the first cycle of a modulation that was evident in 
the high harmonic spectrum [6]. Across a wide range of combinations 
of $(a_0, n_e/n_c)$ Boyd and Ondarza-Rovira found that the frequency 
at which harmonic intensities were modulated was governed by the 
plasma frequency. A Fourier decomposition of the plasma electron 
density showed correspondingly strong modulation at the plasma 
frequency so that harmonics of the laser frequency $\omega_L$ such that 
$m \gtrsim 1.5\, (\omega_p/\omega_L)$ reflect this modulation in 
density, {\it cf.\ }[6]. 

In the UR-regime the laser-plasma interaction physics is strongly 
non-linear. Not only is the plasma line itself broadened, but its 
harmonics may also be strongly driven and likewise broadened. One 
consequence of the non-linearity is some variability in 
the order of plasma harmonics excited for different combinations of 
intensity and electron density. Another effect of emission 
at harmonics of the plasma frequency $l\, \omega_p$ is a widening
of the range of $m$ well beyond the value $m \sim 1.5\, 
(\omega_p/\omega_L)$ at which the first minimum appeared in the 
harmonic spectrum at moderately relativistic energies, Ref.\ 6.
Thus one might now expect the threshold for onset of modulation
to be shifted to higher frequencies, typically $m_{o}^{*} \sim $ 
$(1 - 2)\,l \, (\omega_p/\omega_L)$ where $l$ is 
the highest plasma harmonic excited, which may contribute to 
the variation found in Fig.\ 2.

While Fig.\ 2 shows departures from the universal $p = 8/3$ scaling 
proposed by BGP, fine detail in the structure of these spectra is seen 
more readily over sections of the full harmonic range. In Fig.\ 3a we 
show a Fourier representation of the electron density at a point inside 
the plasma boundary
for the parameters used in Fig.\ 1c. We see from this that the 
structure in the Fourier decomposition of electron density is reflected
in the power spectrum. Above $m \sim 4\, \omega_{p}/\omega_{L} 
\sim 25$ is a region extending to $m \sim 100$ over which the Fourier
amplitudes show only modest decay, a detail 
mirrored in the behaviour of $P_m$ found in Fig.\ 1c, where 
there is a clear discontinuity in the decay index in this region.

As an illustration of the variability referred to earlier, we show 
in Fig.\ 3b a Fourier density representation for the parameter 
combination in 
Fig.\ 2d, i.e.\ $a_0 = 20, n_e/n_c = 100$, using a log-linear plot to 
highlight the periodicity of the density modulations. 
This shows a relatively broad range across which there is comparatively 
little variation in Fourier amplitudes; between $m \sim 80$ and 
$m \sim 140$ we find modulation at the plasma frequency. Above this 
the modulation frequency switches to approximately $2\, \omega_p$. The 
triplet that appears over $170 \lesssim m \lesssim 230$ is the source 
of broadened emission over this range in Fig.\ 2d, typically 
$\sim 6\, \omega_{p}$ wide though this is difficult to
discern given the poor resolution in Fig.\ 2d.

\renewcommand{\thefigure}{3}
\begin{figure}[h]  
{\bf \epsfxsize=9.0cm \epsfysize=9cm \epsffile{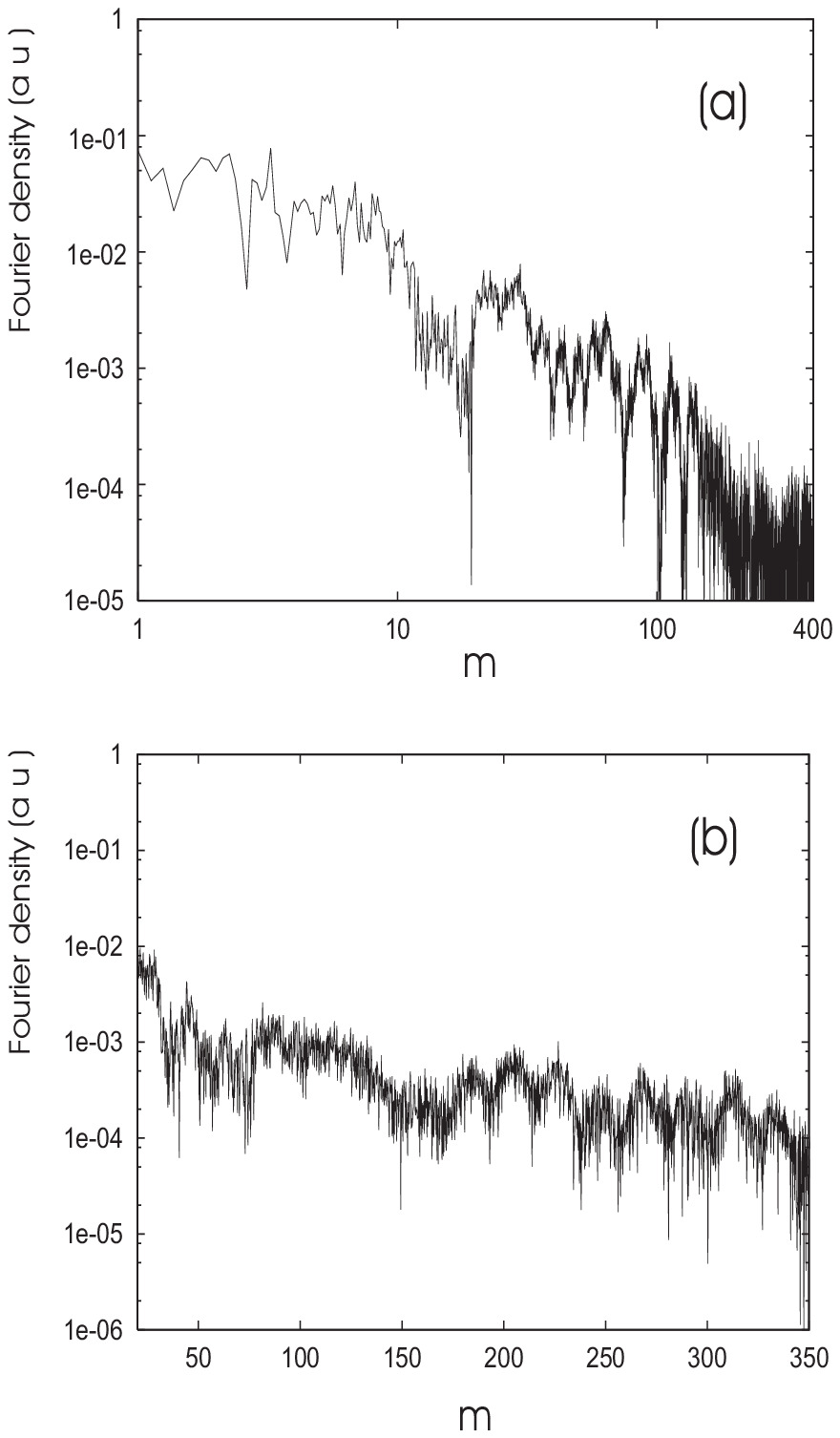}\vspace*{0.5cm}}
\parbox{9cm}{Figure 3: \small (a) Fourier decomposition of the electron 
density corresponding to the parameters used to generate the spectrum 
shown in Fig.\ 1c;
(b) a log-linear representation of Fourier amplitudes across the range 
$20 \leq m \leq 350$ for the harmonic spectrum in Fig.\ 2d, where 
$a_0 = 20, n_e/n_c = 100$.}
\end{figure}

\noindent The range of values of the decay index in the moderately
relativistic range $0.5 \lesssim a_0 \lesssim 2.0$ considered in 
Ref.\ [6] varied from $p \sim 4$ to $p \sim 8/3$, for parameter choices
such that typically $S = n_e/(a_0\, n_c) \gtrsim 20$. 
In the UR regime $S \lesssim 10$ for the 
three $a_0$ values used, $a_0 = 5, 10, 20$. Fig.\ 4 shows the decay 
index from PIC runs spanning this range of parameters. For $a_0 = 5$ 
the decay index was found to be $p = 7/3$, irrespective of density.
For higher $a_0$ this reduces to $p=5/3$, or in a few instances, 
$p = 2$.
For each combination considered, the index appears to be 
distinctly weaker than the universal value proposed by BGP.  
In contrasting the values of $p$ appearing in Fig.\ 4 with the BGP
universal value 8/3 we emphasise that nowise do we attribute even 
quasi-universality to a particular value of $p$, be it $p=5/3$ or 
$7/3$. 

That aside, results from this set of simulations support 
our contention that the decay of intensity with harmonic number 
in the ultra-relativistic regime is 
more complex than allowed by the ``universal'' decay 
index $p = 8/3$. The central tenet of the self-similar approach
is that for constant values of the similarity parameter 
$S = n_e/(a_0\, n_c)$, the laser-plasma dynamics remains similar, since 
$S$-similarity corresponds to a multiplicative transformation group of 
the Maxwell-Vlasov equations. While this is indeed
the case for $s$-polarised or normally incident light, $S$ no 
longer serves to characterise the dynamical system uniquely when 
plasma effects contribute to the interaction physics. 

\renewcommand{\thefigure}{4}
\begin{figure}[h]  
\centering  \leavevmode
{\bf \epsfxsize=7.0cm \epsfysize=6cm \epsffile{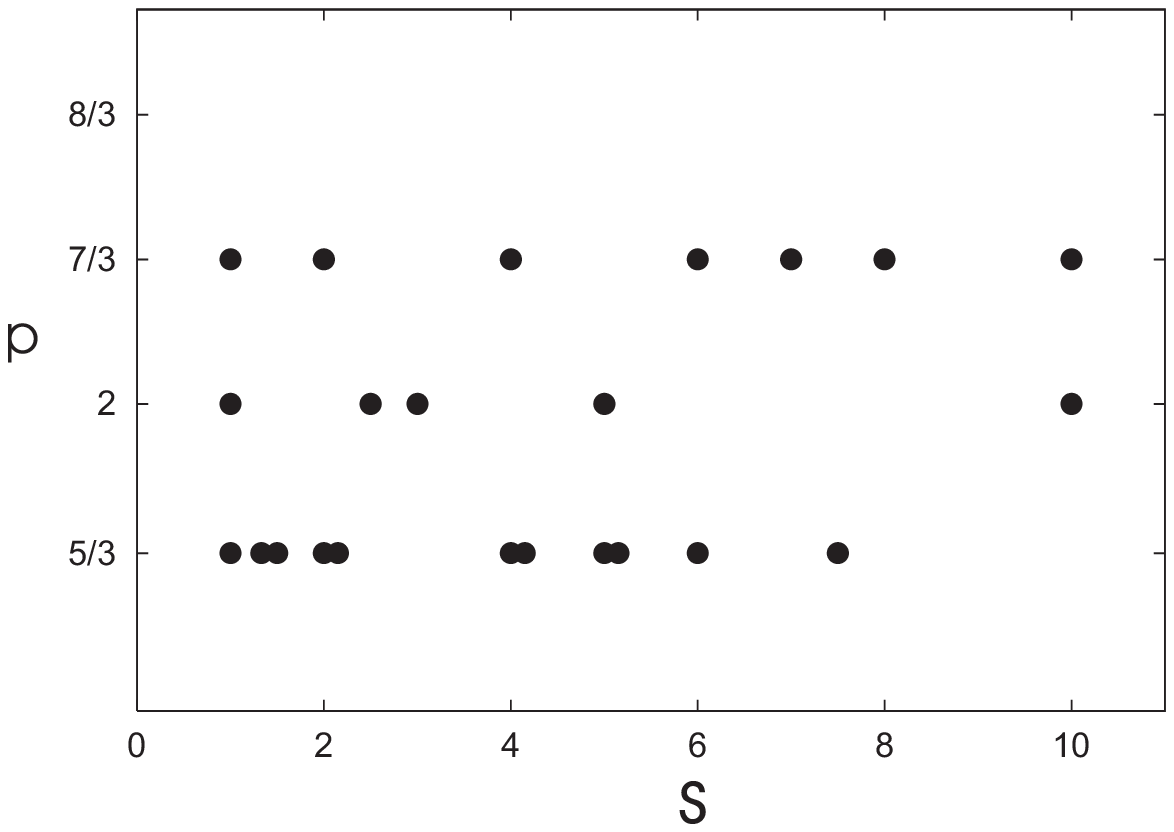} \vspace*{0.5cm}}
\parbox{9cm}{Figure 4: \small Deviations of $p$ from 8/3 universality 
across a range of $S = n_e/(a_0\, n_c)$ for $a_0 = 5, 10, 20$ and a 
range of electron densities. The double solid circles indicate two or 
more data points at $S$= 1.5, 2, 4, and 5.}
\end{figure}

\noindent Thus while the combinations of intensity and electron density
in Figs.\ 1c and 2a correspond to $S = 4$, and those in Figs.\ 2c,d to 
$S = 5$, the interaction physics, at least as far as its footprints in 
the emission spectrum is concerned, does not appear to be similar in 
either case.
Note in passing that by {\it similar} we intend formal $S-$similarity,
as against the conventional usage; in that sense all the spectra are 
similar.
Whereas for the combination in Fig.\ 1c, $p = 5/3$, for Fig.\ 2a, 
$p = 7/3$; similarly $p = 5/3$ for Fig.\ 2c as opposed to $p = 2$ in 
Fig.\ 2d. We conclude that plasma-enhanced emission together with 
plasma-induced modulation can result in decay of the harmonic spectrum 
significantly weaker than that predicted by similarity theory. 
At the highest intensity modelled in this work ($a_0=20$) we have not 
found any $p < 5/3$; that notwithstanding, we do not rule out yet 
lower values at higher intensities. Indeed under more extreme UR 
conditions we have obtained PIC spectra for which $p \sim 4/3$.

One of us (R.O.R.) acknowledges support from CONACyT under Contract 
No.\ 43621-F.

\

\noindent {\bf References}

\noindent [1] N.H.\ Burnett {\it et al.,} Appl.\ Phys.\ Lett.\ 
{\bf 31}, 172 (1977).

\noindent [2] R.L.\ Carman, C.K.\ Rhodes, and R.F.\ Benjamin, Phys.\
\hspace*{4mm} Rev.\ A {\bf 24}, 2649 (1981).

\noindent [3] B.\ Dromey {\it et al.,} Phys.\ Rev.\ Lett.\ {\bf 99}, 
085001 (2007); \hspace*{4mm} Nature Phys.\ {\bf 2}, 456 (2006).

\noindent [4] T.\ Baeva, S.\ Gordienko, and A.\ Pukhov, Phys.\ Rev.\ 
\hspace*{4mm} E {\bf 74}, 046404 (2006).

\noindent [5] T.J.M.\ Boyd and R.\ Ondarza-Rovira, 
Phys.\ Rev.\ \hspace*{4mm} Lett.\ {\bf 85}, 1440 (2000).

\noindent [6] T.J.M.\ Boyd and R.\ Ondarza-Rovira, Phys.\ Rev.\
\hspace*{4mm} Lett.\ {\bf 98}, 105001 (2007).

\noindent [7] U.\ Teubner {\it et al.,} Phys.\ Rev.\ Lett.\ {\bf 92},
185001 (2004).

\noindent [8] K.\ Eidmann {\it et al.,} Phys.\ Rev.\ E {\bf 72},
036413 (2005).

\noindent [9] A.\ Bourdier, Phys.\ Fluids {\bf 26}, 1804 (1983).

\noindent [10] F.\ Brunel, Phys.\ Rev.\ Lett.\ {\bf 59}, 52 (1987).

\end{document}